# Colorimetry and Tribology of Ultrapure Copper Surface Micromodification

Aleksandra Szczupak[1], Grzegorz Cios[2], Benedykt R. Jany[1*]

[1]Marian Smoluchowski Institute of Physics, Faculty of Physics, Astronomy and Applied Computer Science, Jagiellonian University, ul. prof. Stanisława Lojasiewicza 11, 30-348 Krakow, Poland
[2]Academic Centre for Materials and Nanotechnology, AGH University of Krakow, al. A Mickiewicza 30, 30-059 Krakow, Poland
*corresponding author e-mail: benedykt.jany@uj.edu.pl

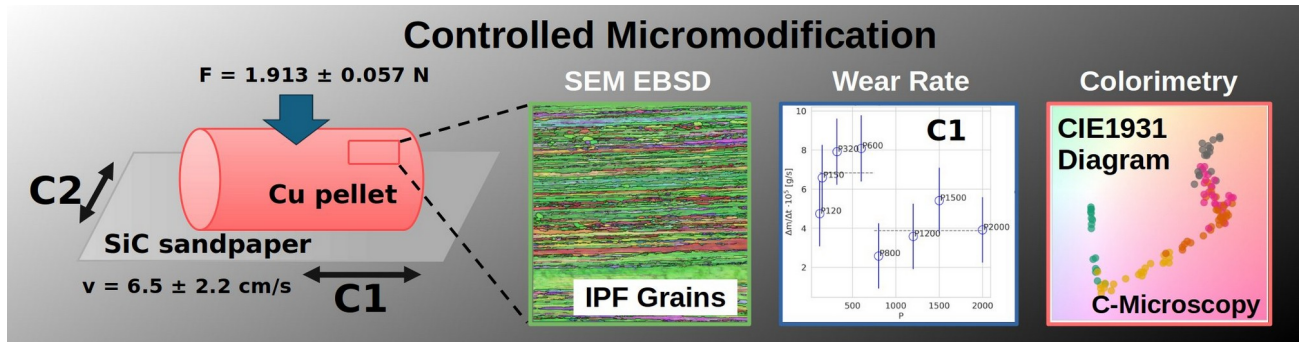

## Abstract

Controlling optical and tribological properties of metal surfaces, like color and wear rate, without altering their chemical composition is a highly desirable process across numerous fields of science and industry. It represents a cost-effective alternative to traditional chemical methods, particularly for copper, one of the most important metals widely used where high electrical and thermal conductivity, alongside resistance to corrosion, are required. We investigated the control of copper surface texture through a controlled micromodification process, utilizing constant force and velocity with abrasive silicon carbide sandpaper on ultrapure copper pellets exhibiting elongated crystallographic grains, and its impact on optical properties. Systematically varying grit size and rubbing direction, both along and across the grains, resulted in tunable microgroove morphology, demonstrating a marked difference in wear rate between single-grain and multi-grain abrasion. Furthermore, modification along copper grain boundaries yielded a change in the wear rate by a factor of two, related to single-grain and multi-grain abrasion regime changes, enabling precise control over material performance via tuned abrasion conditions. Colorimetric analysis via C-Microscopy revealed a strong, statistically significant relationship between abrasive parameters, microgroove geometry (inclination angle, depth, and size), and optical spectral signatures, which were then parametrized to achieve targeted control. This research demonstrates a simple yet effective approach to color and reflectance modification via microgroove engineering, offering a pathway to customized material properties by uniquely coupling contact mechanics, surface morphology, and colorimetry at the microscale level.

**Introduction**

Copper (symbol Cu) is a chemical element with an atomic number of 29 belonging to group 11, period 4, and the d block of the periodic table of chemical elements. It is a reddish-brown, malleable metal with very high thermal conductivity (401 Wm$^{-1}$K$^{-1}$ ) and electrical conductivity (56.9 MSs$^{-1}$), characterized by a distinctive orange-red color[1]. Copper, both pure and in the form of alloys with other metals, is widely used in all areas where high electrical and thermal conductivity and resistance to corrosion are required[2]: in electronics, energy, and telecommunications for the production of electrical wires[3], transformers, electromagnets, or micro-scale systems[4], as well as in transportation and construction[5], biology[6] and medicine[7], as well as in art and decoration[8]. Most metals are gray or silvery-white. Only a few metals, such as copper, gold, cesium, or osmium, are characterized by exhibiting other colors determined by the structures of their valence bands. The orange-red color of copper is caused by transitions of electrons from the 3d electron orbital to a partially filled 4s shell[9]. Altering optical properties in particular color of metal surfaces without changing their chemical composition is a highly desirable process in many fields of science, industry, or art. It represents a cost-effective alternative to traditional chemical methods. It is also an environmentally friendly technique. Many different approaches are used, like laser-induced metal surface coloring, which uses laser to create color changes through localized surface modifications[10], formed surface structures reflect specific wavelengths of light, producing a visible color. Another example is the formation of thin oxide layers causing interference effects, resulting in changes in the reflectance spectrum (i.e. color) of the metal surface[11]. As recently shown, structural coloring of surfaces can be also obtained based on the formation of microscale morphological structures causing diverse optical effects linked to light interaction with rough surfaces[12]. The studies showed that it is possible to explain the observed colors of metals by interacting with light and their surface morphological features: roughness and the distribution of various structures, grooves, ledges, and crevices, where light is multiply reflected according to the laws of optics.

Interestingly, metal color changes, especially noticeable discolorations, are key indicators of tribochemical reactions occurring during friction and wear[13]. These changes reveal the formation of tribofilms – thin films created on contacting surfaces – and signal shifts in the material's surface chemistry and structure. The tribology of metals, in particular copper, plays a critical role in improving material selection and system design to minimize energy losses, boost durability, and enhance the efficiency of mechanical components such as engines and gears[14,15,16]. While metals are widely employed for their favorable tribological properties, accurately understanding their wear behavior is essential for predicting component lifespan and extending service life[17,18,19,20]. Excessive wear results in wasted energy, costly replacements, and higher emissions - underscoring the need for advanced wear-resistant materials and innovative technologies.

The ability to precisely control the optical properties of materials and its wear rate is crucial for a wide range of applications. We investigated the control of copper surface texture through a controlled micromodification process, utilizing constant force and velocity with abrasive silicon carbide sandpaper on ultrapure copper pellets exhibiting elongated crystallographic grains, a characteristic beneficial for enhanced grain boundary modification. Systematically varying grit size and rubbing direction, both along and across the grains, resulted in tunable microgroove morphology, demonstrating a marked difference in wear rate between single-grain and multi-grain abrasion. Furthermore, modification along copper grain boundaries yielded a change in the wear rate by a factor of two, related to single-grain and multi-grain abrasion regime changes. Colorimetric analysis via C-Microscopy revealed a strong, statistically significant relationship between abrasive parameters, microgroove geometry (inclination angle, depth, and size), and optical signatures, which were then used to achieve targeted microgroove geometry. Validation via a multiple reflections model confirmed the driving force behind color and reflectance changes. This research lays the groundwork for developing novel surface modification techniques with precise control over optical properties.

**Materials and Methods**

In the controlled micromodification experiment, 18 samples of ultra-pure copper Cu 99.999% were used in the form of a pellet (Kurt J. Lesker Company) measuring 1/8 inch in length and 1/8 inch in diameter, with a mass of 0.213 ± 0.001 g. Each sample was subjected to controlled micro-modification using silicon carbide (SiC) abrasive paper with a specified grain size (P). The sample was uniformly pressed against the abrasive paper with a force F = 1.913 ± 0.057 N, and then rubbed in the direction along C1 or across C2 of the crystallographic grains at a constant velocity of v = 6.5 ± 2.2 cm/s for a duration of 60 seconds. Subsequently, each sample was cleaned separately in isopropanol using an ultrasonic washer for 300 seconds. To determine the applied force F during modification, an electronic weighing scale WPT 3C No. 127783/04 RADWAG acting as a force sensor was used. From the images of the trajectories formed on the surfaces of the abrasive papers, the velocity of modification v was determined by measuring the formed trajectories by ImageJ/FIJI[21] program. To characterize the abrasive papers used the distance between the sandpaper grains $o_p$ and sandpaper grain diameter $d_d$ were calculated from the collected sandpaper micrographs by ImageJ/FIJI program. To investigate the mass loss of the copper pellets during micro-modification, wear – an electronic precise weighing scale AS220/X No. 358170/12 RADWAG was used, with which the masses of the original and modified samples were measured, and subsequently the ratio of their differences and the time of micro-modification $\Delta m/\Delta t$ was calculated, where $\Delta t$ = 60 seconds is the modification time.

The surface of the modified samples was investigated by colorimetric microscopy (C-Microscopy)[22], the Delta Optical Smart 5MP PRO digital microscope was used which was color calibrated using Calibrite ColorChecker Classic Mini calibration card and D65 illuminant accordingly. The local information (from microregion) about the surface texture and optical properties: color in chromaticity coordinates CIE1931 and reflectance R were measured.

The microstructure, in particular crystallographic grains, were measured by SEM Versa 3D (Thermo Fischer Scientific, Waltham, MA, USA) equipped with a Symmetry S2 EBSD detector (Oxford Instruments Nanoanalysis, High Wycombe, UK). On a lateral surface of the copper pellet. The map was collected at 20 kV and ~20 nA with diffraction pattern resolution of 156x128, and step size of 500 nm. For pattern indexing standard Hough based approach indexing implemented in Aztec v. 6.1 software was used and Aztec Crystal v. 3.1 to analyze the data and generate the images. The sample for EBSD investigation was polished with IM4000Plus ion milling system.

**Results and Discussion**

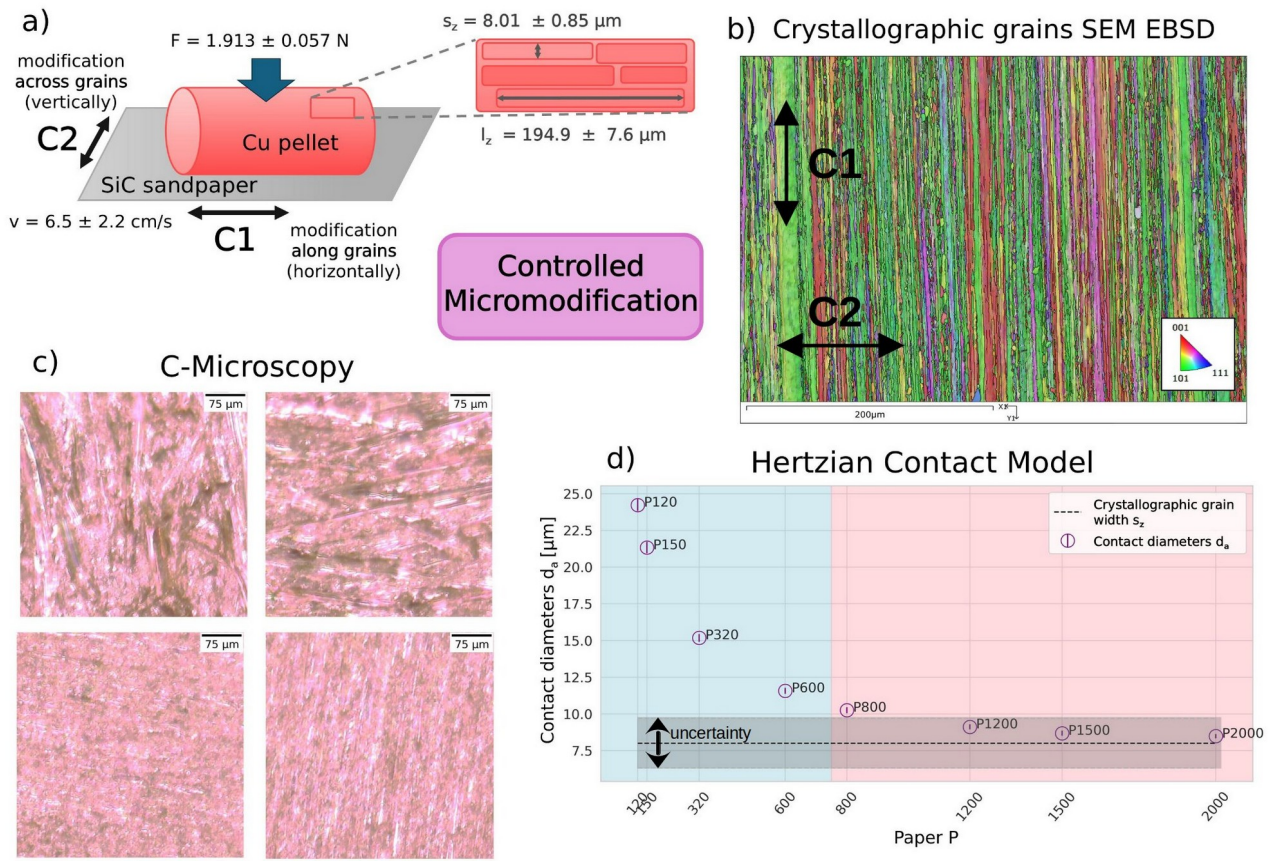

*Figure 1: a) Scheme illustrating controlled micromodification experiments performed on ultrapure copper samples. The micromodification experiments were performed on copper pellets with a constant Force (F) and velocity (v) b) SEM EBSD Inverse Pole Figures (z out of plane) revealing elongated grains in a copper pellet. c) Example C-Microscopy images, colorimetrically calibrated under a D65 illuminant, of copper surfaces following modification. d) Comparison of the contact diameters of abrasive paper grains on the copper surface, determined using Hertzian Contact Model, as a function of sandpaper grit (SiC grain size). The crystallographic grain width with uncertainty is indicated. Two distinct regimes are visible: one where the contact diameter exceeds the crystallographic grain width, and the second where the contact diameter is the same as the crystallographic grain width.*

Figure 1a) shows a scheme illustrating controlled micromodification experiments performed on ultrapure copper samples. The used copper pellet samples exhibited elongated crystallographic grains, as revealed by SEM-EBSD Inverse Pole Figures (z out of plane), Fig. 1b). These grains had an average width of $s_z$ = 8.01 ± 0.85 μm and an average length of $l_z$ = 194.9 ± 7.6 μm (see also Fig. S1 in Supporting Information). The EBSD data further indicated that the dominant copper surface crystallographic plane is (101), as shown in the Pole Figure and Inverse Pole Figures (Fig. S2-S3 in Supporting Information). Controlled micromodification was achieved by uniformly pressing the copper pellets against abrasive silicon carbide (SiC) sandpaper with varying grit grades (P120-P2000 i.e., different grain sizes) using a constant force F = 1.913 ± 0.057 N. The pellets were then rubbed along C1 or across C2 of the crystallographic grains at a constant velocity of v = 6.5 ± 2.2 cm/s for 60 seconds. The experiments resulted in the formation of different surface textures, as revealed by C-Microscopy (Fig. 1c), characterized by microgrooves. To better understand the experimental process, the Hertzian contact model was used to calculate the real contact diameter $d_a$ between the SiC grain and the copper. The results are presented in Fig. 1d). It was observed that increasing the SiC grit grade causes a decrease in contact diameter – details are presented in Table S1 in Supporting Information. Plotting the copper crystallographic grain width together with uncertainty ($s_z$, black dashed line, Fig. 1d)) reveals two distinct regimes: one where the contact diameter exceeds the crystallographic grain width (P120 to P600), and the second where the contact diameter is the same as the crystallographic grain width (P800 to P2000).

To describe the formed surface texture, characterized by microgrooves, a 2D autocorrelation analysis was performed on the C-Microscopy images of the surface Fig. 2a). For each image, autocorrelation parameters (characteristic distance $d_{ch}$ and structure size $s_s$) were extracted in the vertical (V) and horizontal (H) directions, as indicated. Detailed roughness analysis was performed Fig. 2b). Surface roughness was extracted from cross-sectional measurements using image analysis with ImageJ/FIJI. Details are presented in Fig. S5 in Supporting Information; such an analysis is equivalent to Atomic Force Microscopy (AFM) or stylus profilometry measurements[23]. A full height profile was extracted, subsequently decomposed into a high-frequency component (roughness) and a low-frequency (waviness) component. The high-frequency cutoff parameter was set to 80 μm, according to ISO 14238-1:2008. The analysis revealed statistically significant correlations (p-value < 0.05) between sandpaper parameters and texture parameters, as shown in Fig. 2c). Specifically, correlations were observed between the distance between sandpaper grains ($o_p$) and the characteristic distance from autocorrelation ($d_{ch}$) for: the modification C1, in the horizontal (H) direction; and the modification C2, in the vertical (V) direction. Furthermore, correlations were found between the sandpaper contact diameter ($d_a$) and the structure size from autocorrelation ($s_s$) for: the modification C1, in the horizontal (H) direction; and the modification C2, in the vertical (V) direction. Additionally, a statistically significant correlation was observed between the depth of maximum shear stress in copper ($z_{max}$, from the Hertzian model) and the RMS roughness ($R_q$) for the modification C1. This is a particularly interesting result, as it allows for the prediction of surface roughness through Hertzian model calculations. Finally, a correlation was detected between the distance between the sandpaper grains and the RMS waviness ($W_q$) for the modification C1. The analysis also identified further statistically significant correlations between sandpaper parameters and surface structures, as depicted in Fig. S6-S11 in Supporting Information. The analysis shows that controlling the sandpaper parameters allows control over the parameters of the surface texture.

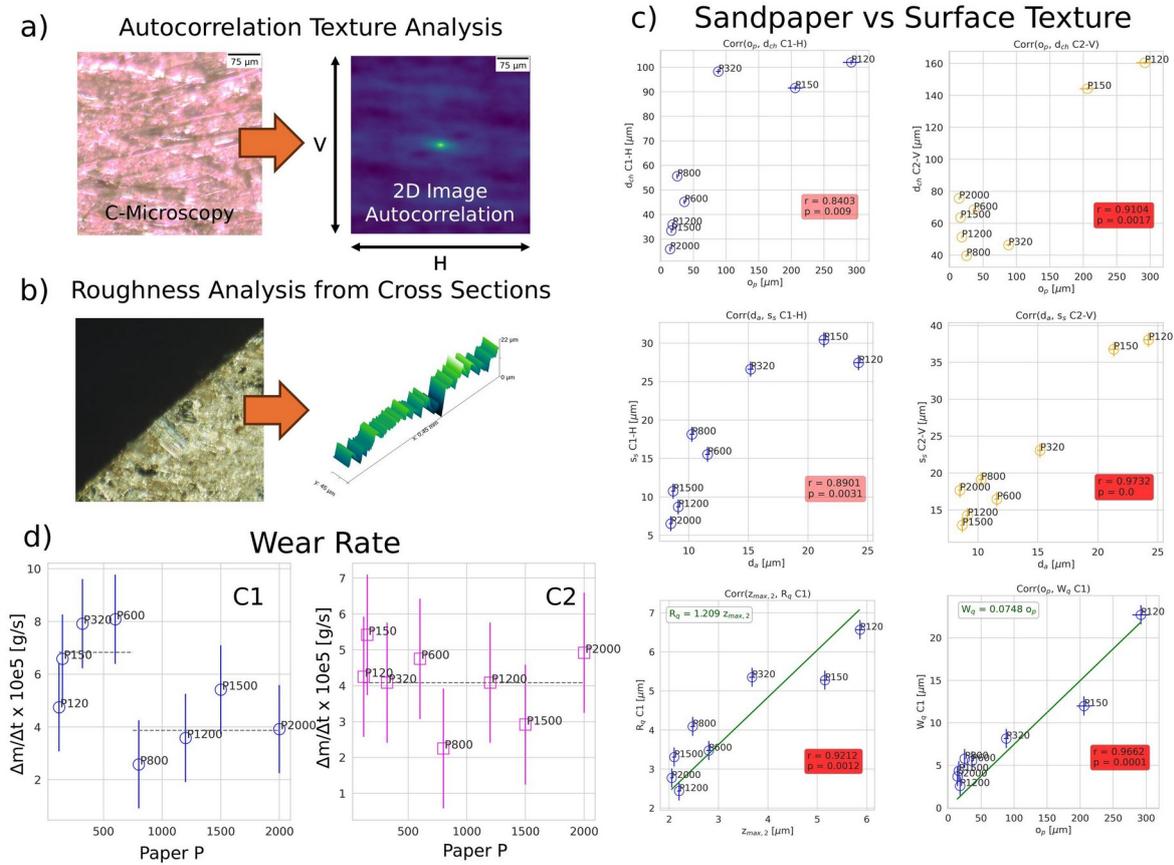

*Figure 2: a) Scheme illustrating 2D autocorrelation texture analysis derived from C-Microscopy colorimetrically calibrated (D65 illuminant) images of modified samples. The autocorrelation parameters are measured in the direction vertical V and horizontal H, as indicated. b) Scheme illustrating roughness analysis performed on samples' cross-sections. c) Relevant correlations between sandpaper parameters and texture parameters. First row, correlation between distance between the sandpaper grains $o_p$ and characteristic distance from autocorrelation $d_{ch}$ for: the modification C1 and direction H; the modification C2 and direction V. Second row, correlation between sandpaper contact diameter $d_a$ and structure size from autocorrelation $s_s$ for: the modification C1 and direction H; modification C2 and direction V. Last row, Correlation between depth of maximum shear stress in copper $z_{max}$ (from Hertzian model) and roughness $R_q$ for the modification C1. Correlation between distance between the sandpaper grains and waviness $W_q$ for the modification C1. The correlation coefficient and p-value is given in the red box. It is seen that by controlling the sandpapaer parameters one can control the parameters of the surface texture. d) Wear rate during the abrasion process of modified samples as a function of sandpaper grit (SiC grain size), for the modification along – C1 and across – C2 crystallographic grains. It's observed that the wear rate for modification along the grains (C1) changes by approximately a factor of two with change in sandpaper grain size.*

Finally, a very interesting behavior of the wear rate on sandpaper grit (SiC grain size) was discovered Fig. 2d). For the modification direction along the crystallographic grains (C1), the wear rate exhibits two distinct regimes – a high wear rate and a low wear rate, differing by approximately a factor of two. This coincidence aligns precisely with two distinct contact diameter regimes Fig. 1d). When the process occurs along the crystallographic grains, abrasion takes place within several crystallographic grains and their boundaries, resulting in a higher wear rate. Conversely, when the contact diameter equals the crystallographic grain width, abrasion occurs within a single crystallographic grain, yielding a lower wear rate. However, when the modification direction is across the crystallographic grains (C2), no such effects are observed; the wear rate remains constant. This is because the abrasion process consistently occurs within the same manner, involving many grains and boundaries in this direction.

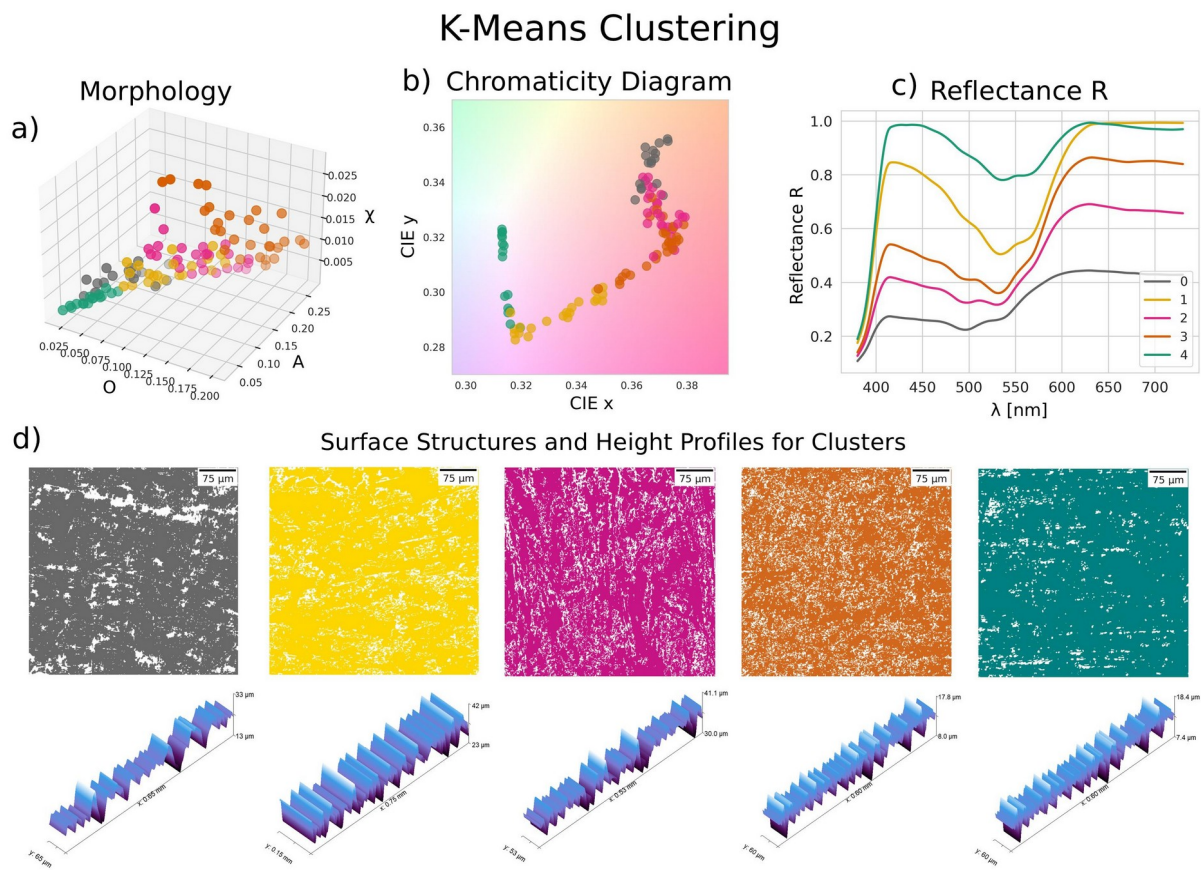

*Figure 3: a) Grouping/clustering of copper surface morphologies after modification using K-Means. Minkowski functionals representing the 109 obtained surface structures: A represents area, O represents perimeter, and χ represents the Euler characteristic (5 clusters indicated by different colors). b) Optical properties, as described by chromatic coordinates on the CIE 1931 chromaticity diagram, grouped for the 109 obtained surface structures (5 clusters indicated by different colors). c) Mean cluster reflectances R as a function of wavelength. d) Representative mean cluster surface structures and microgroove height profiles for 5 clusters.*

Earlier it was observed that there is an effect of a friction changes on the grain direction alignment[24] and on the grain boundaries[25]. The found wear rate behavior has a numerous different applications in materials processing and industry applications, since by changing only the abrasion medium parameters to fit the material's grain structure, one can control the wear rate behavior of the material.

For all 18 measured surfaces, a hyperspectral reflectance (R) was reconstructed and subsequently grouped/clustered using K-Means clustering, resulting in surface regions (morphologies) with similar spectral reflectance, as detailed in B.R. Jany[22], and further illustrated in Fig. S12 in Supporting Information. This process identified in total 109 distinct surface morphologies made of surface microgrooves. To characterize these morphologies, we utilized Minkowski Functionals[26],[27] (Fig. S13 in Supporting Information): A represents area, O represents perimeter, and χ represents the Euler characteristic, all normalized to the total number of pixels on the image, see Fig. 3a). We also calculated optical parameters, specifically the chromaticity coordinates (CIE x and CIE y), for each morphology Fig. 3b). Analysis reveals that the morphologies of these surface structures span across the entire space of the Minkowski Functionals, and exhibit distinct, non-random shapes within the chromaticity diagram, suggesting the presence of highly specific microgroove geometries. We also observed a statistically significant correlation between the morphological and optical properties, see Fig. S14 in Supporting Information. This prompted us to cluster/group all of the optical and morphological data for the 109 distinct morphologies. We again utilized K-Means clustering, evaluating the optimal number of clusters/groups using information criteria AIC and BIC, see Fig. S15 in Supporting Information. This resulted in the data being grouped into five optimal clusters indicated by different colors in Fig. 3. We also calculated the mean cluster reflectance (R) as a function of wavelength Fig. 3c), revealing that each cluster exhibits a distinct reflectance profile and a unique spectral signature. Furthermore, we observed a particularly strong and consistent monotonic relationship between the Minkowski Functionals and the mean reflectance for each cluster, see Fig. S16 in Supporting Information, indicating a connection between surface geometry and optical properties. This relationship, coupled with the identification of five distinct clusters, suggests a clear underlying organization of surface morphologies based on both geometrical and spectral characteristics. This detailed analysis of five clusters provides valuable insights into the relationship between surface structure and optical response. Finally, Fig. 3d) illustrates representative mean cluster surface structures and microgroove height profiles, obtained by filtering the original height profiles based on the size of the corresponding mean cluster surface structures, for each of the five identified clusters.

Based on the observed surface structures and height profiles for each cluster, we evaluated the parameters of the formed microgrooves. These microgrooves can be characterized by their inclination angle – defined as the angle between the z-direction and the normal to the facet Fig. 4a) – alongside microgroove depth ($R_{vm}$) and size. Microgroove size was calculated as the average dimension of the structures, defined as the major axis of a fitted ellipse using the ImageJ/FIJI program. Microgroove depth ($R_{vm}$) was calculated as the average maximum valley depth from the height profiles using the Gwyddion[28] software. Considering the geometry of these microgrooves, with depth and size, allows for the calculation of their angle Fig. 4b).

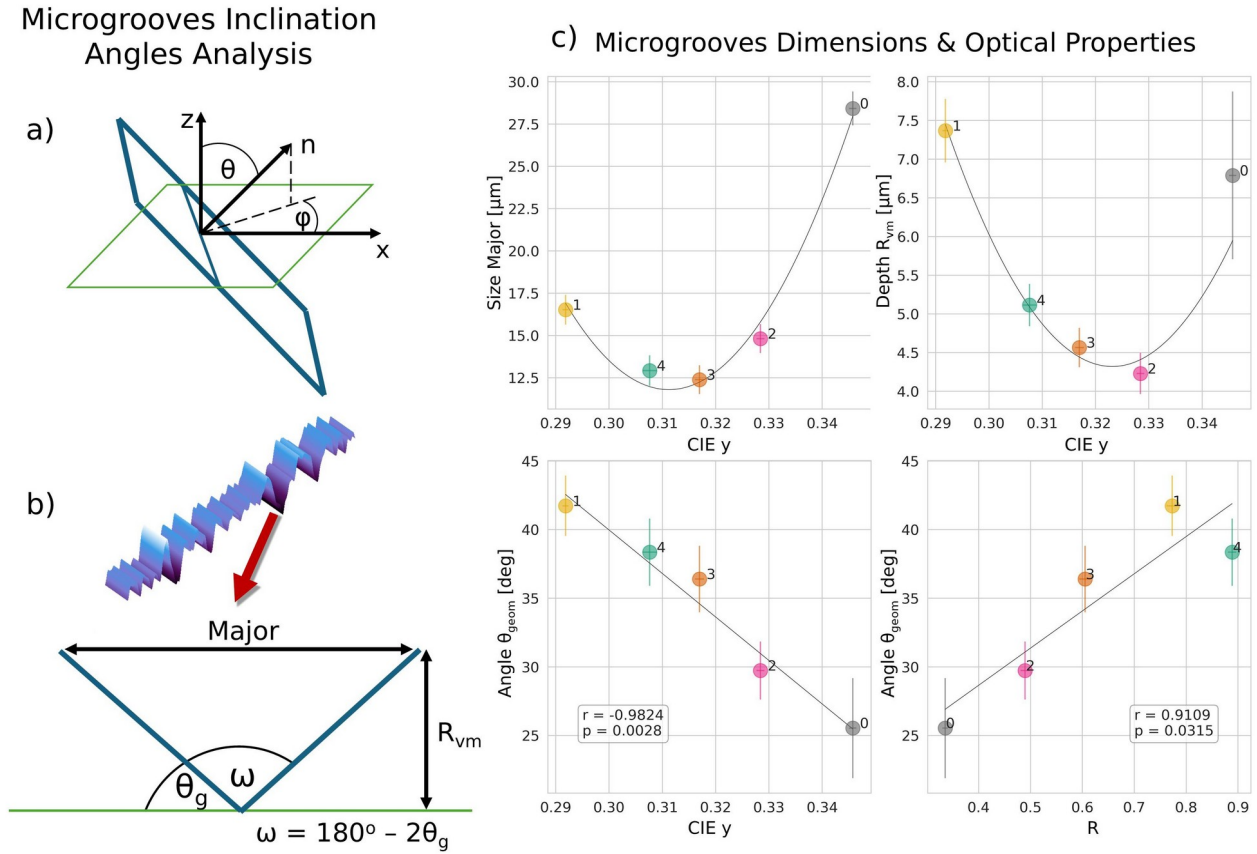

Figure 4: a) Microgrooves Inclination Angles Analysis. Definition of the inclination angle. b) Geometricaly calculated inclination angle θgeom. c) Relations between optical parameters: chromatic coordinates CIE y, reflectance R and morphological parameters: mean maximum width of microgrooves Major Size, mean microgrooves depth Rvm, mean maximum inclination angle θgeom of average cluster structures of the formed microgrooves of the copper surface. It is seen that by controlling the formed microgrooves parameters one can control local optical properties (chromaticities and reflectance).

Fig. 4c) shows the dependence of microgroove dimensions on the optical properties. It is observed that as the size and depth of the microgrooves change, the chromaticity coordinate CIE y changes monotonically. This relationship can be parameterized by a quadratic function: $y = ax^2 + bx + c$. Specifically, for size, we obtain a = (13600 ± 1300) µm, b = (-8500 ± 810) µm, and c = (1330 ± 130) µm; and for depth, a = (3190 ± 740) µm, b = (-2060 ± 470) µm, and c = (337 ± 73) µm. We also observe a linear relationship between the microgroove angle on CIE y and on reflectance R, which was parameterized by the function y = ax + b. The parameters for CIE y are: a = (-317 ± 66) deg and b = (135 ± 21) deg, while for reflectance R, the parameters are a = (27.1 ± 6.2) deg and b = (17.8 ± 4.2) deg. Changing the geometry of the microgrooves allows for a change in color, specifically as represented by the chromaticity coordinate CIE y. It is also noteworthy that as the microgroove angle increases, the reflectance also increases linearly, providing a unique opportunity to control the surface reflectance through specific microgroove fabrication.

The derived microgroove parameters were used to calculate spectral reflectance via multiple reflections, driving the observed changes in color and reflectance, as proposed by A. Manas[12]. A comparison between the model and the experimental data is presented in Fig. S17-S18 in the Supporting Information. The model reasonably reproduces the main measured reflectances, although it exhibits some discrepancies in the low wavelength region (400-500 nm). These discrepancies, with a range of approximately 5% to 30%, are systematic and correlate with changes in microgroove angle, see Fig. S19 in Supporting Information. These limitations may be attributed to factors not included in the model, such as subfacceting of the microgroove facets, plasmonic effects, or asymmetric facets. Nevertheless, the multiple reflections model, on average, reproduces the collected experimental data.

All experimental data, including raw measurements and databases of measured properties, are freely available in the Zenodo repository[29].

**Summary and Conclusions**

This study systematically investigated the control of surface texture on copper through a controlled micromodification process, with constant force and velocity, utilizing abrasive silicon carbide sandpaper and its impact on the optical properties. Controlled variation in the grit size of the sandpaper, combined with specific rubbing directions, on ultrapure copper pellet with elongated crystallographic grains, (along copper grain boundaries, across copper grain boundaries), resulted in a highly tunable surface morphology characterized by microgrooves. Importantly, the observed wear rate exhibited a strong dependence on the applied rubbing direction and the contact diameter, showing a marked difference between abrasion occurring within a single grain versus across multiple grains. For the modification along copper grain boundaries, this manifests as a change in the wear rate by a factor of two, enabling precise control over material performance through tailored abrasion conditions. Colorimetric and reflectance analysis by C-Microscopy measurements, revealed a complex and statistically significant connection between the abrasive parameters, the resulting microgroove geometry (including inclination angle, depth, and size), and the surface's optical properties. The data clustering by K-Means identified distinct surface morphologies, each possessing unique spectral signatures and corresponding Minkowski Functionals, showcasing a high degree of surface heterogeneity. Furthermore, the microgrooves were found to be intrinsically linked to color, as evidenced by a monotonic relationship between microgroove size/depth and the CIE y chromaticity coordinate. The linear dependence was discovered between microgroove angle and a reflectance R. All these relations were successfully parametrized. This together effectively allows for surface color and reflectance R changes through geometric control. The obtained spectral reflectance data were also successfully compared with reflectance calculation via multiple reflections, driving force of the observed changes in color and reflectance.

Ultimately, this research demonstrates a powerful and simple approach for tailoring the surface properties like color and reflectance of copper by particular microgroove engineering. This control is rooted in the direct coupling of contact mechanics, surface morphology and colorimetry at microscale level providing a significant advancement in surface engineering techniques and offering a clear pathway for customized material properties with applications across various industries.

**Acknowledgments**

This research was supported in part by the Excellence Initiative - Research University Program at the Jagiellonian University in Krakow. We would like to acknowledge Enrico Gnecco, Nicola Manini, Reberto Guerra for fruitful discussion related to tribology. This work is the result of Aleksandra Szczupak master thesis.

**Author Contributions**

A.S. performed all copper micromodification experiments together with colorimetric microscopy (C-Microscopy) measurements and data analysis. G.C. contributed to SEM EBSD measurements and data analysis. B.R.J. conceived the idea, supervised and organized the project. B.R.J. prepared the manuscript in consultation with all authors. All authors contributed to the discussion and interpretation of the final results.

**Data Availability**

All the experimental collected data are freely available in Zenodo repository https://doi.org/10.5281/zenodo.17287173.

# References

1 Lide, D.R. (ed.). CRC Handbook of Chemistry and Physics. CRC Press, 2004.
2 Emsley, J. Nature's building blocks: an A-Z guide to the elements. Oxford University Press, 2003
3 Mao, Q.; Zhang, Y.; Guo, Y. et al. "Enhanced electrical conductivity and mechanical properties in thermally stable fine-grained copper wire." Commun. Mater., vol. 2, 2021, p. 46, https://doi.org/10.1038/s43246-021-00150-1
4 Y.Q. Chang, L.W. Kong, X.L. Zhu, X.F. Zhu, J. Cao, B. Wen, P. Li. "Investigation on the strengthening behaviour of micro-scale copper fiber." Materials Science and Engineering: A, vol. 859, 2022, p. 144186, https://doi.org/10.1016/j.msea.2022.144186
5 Aman Deep, Pradip Sarkar. "Enhancing sustainability in concrete construction: Utilizing copper slag for improved properties of geopolymer concrete." Construction and Building Materials, vol. 453, 2024, p. 139044, https://doi.org/10.1016/j.conbuildmat.2024.139044
6 Festa, R. A.; Thiele, D. J. "Copper: An essential metal in biology." Current Biology, vol. 21, no. 21, 2011, pp. R877–R883, https://doi.org/10.1016/j.cub.2011.09.040
7 Peter Tsvetkov et al. "Copper induces cell death by targeting lipoylated TCA cycle proteins." Science, vol. 375, 2022, pp. 1254–1261, https://doi.org/10.1126/science.abf0529
8 Eggert, G. "Copper and bronze in art and the search for rare corrosion products." Heritage, vol. 6, 2023, pp. 1768–1784, https://doi.org/10.3390/heritage6020094
9 Trigg, G.L. (ed.). Encyclopedia of Applied Physics. VCH Publishers, 1992
10 Qin, X.; Xue, Z.; Wang, X.; Song, K.; Wan, X. "A color reproduction method for exploring the laser-induced color gamut on stainless steel surfaces based on a genetic algorithm." Appl. Sci., vol. 15, 2025, p. 28, https://doi.org/10.3390/app15010028
11 Groussin, B., et al. "Efficient Composite Colorization of Copper by Spatially Controlled Oxidation with Deep-UV Ultrafast Lasers." Advanced Optical Materials, vol. 12, 2024, pp. 2302071, https://doi.org/10.1002/adom.202302071
12 Manas, A. "Gold's red shift: colorimetry of multiple reflections in grooves." Gold Bulletin, vol. 53, 2020, pp. 147–158, https://doi.org/10.1007/s13404-020-00285-y
13 Zhu, Y. and Li, H. and Li, F. and Zhu, D. and Ji, L. and Liu, X. and Zhou, H. and Chen, J. "Relationship between discoloration, condensation, and tribology behavior of steel surfaces at cryogenics in vacuum." Tribology International, vol. 191, 2024, pp. 109104, https://doi.org/10.1016/j.triboint.2023.109104
14 Guoqing Zhang, Jun Tang, Kang Yang, Ruili Wang, Yang Chen, Yahui Xiong, Chao Wu, Zhenjie Li, Yaqiong Wang, Haibo Lin. "Important contributions of metal interfaces on their tribological performances: From influencing factors to wear mechanisms." Composite Structures, vol. 337, 2024, p. 118027, https://doi.org/10.1016/j.compstruct.2024.118027
15 Deshpande, P.K.; Lin, R.Y. "Wear resistance of WC particle reinforced copper matrix composites and the effect of porosity." Mater. Sci. Eng., A, vol. 418, 2006, p. 137, https://doi.org/10.1016/j.msea.2005.11.036
16 Bower, A.F. "Cyclic hardening properties of hard-drawn copper and rail steel." J. Mech. Phys. Solids, vol. 37, 1989, p. 455, https://doi.org/10.1016/0022-5096(89)90024-0
17 T.S. Eyre. "Wear characteristics of metals." Tribology International, vol. 9, no. 5, 1976, pp. 203–212, https://doi.org/10.1016/0301-679X(76)90077-3
18 M. Scherge, D. Shakhvorostov, K. Pöhlmann. "Fundamental wear mechanism of metals." Wear, vol. 255, no. 1–6, 2003, pp. 395–400, https://doi.org/10.1016/S0043-1648(03)00273-4
19 Chen, K. and Wu, X. and Zhang, A. and Zhang, J. and Chen, X. and Zhu, Y. and Wang, Z. "Development of wear resistant Cu-12Sn-1.5Ni alloy via minor addition of Fe during casting process." Applied Surface Science, vol. 573, 2022, pp. 151623, https://doi.org/10.1016/j.apsusc.2021.151623
20 Yu, H. and Liang, W. and Miao, Q. and Yin, M. and Jin, H. and Sun, Y. and Xu, Y. and Chang, X. "Study on the wear resistance and thermodynamic stability of (MNbTaZrTi)N (M = Cr, Hf) high-entropy nitride coatings at elevated temperatures." Surface and Coatings Technology, vol. 497, 2025, pp. 131792, https://doi.org/10.1016/j.surfcoat.2025.131792
21 Schindelin, J., et al. "Fiji: an open-source platform for biological-image analysis." Nature Methods, vol. 9, no. 11, 2012, pp. 676–682, https://doi.org/10.1038/nmeth.2019
22 Jany, B.R., "Quantifying Colors at Micron Scale Using (C-Microscopy)." Microscopy, vol. 176, 2024, 103557, https://doi.org/10.1016/j.micron.2023.103557
23 Balachandran, S., et al. "A Method for Measuring Interface Roughness from Cross-Sectional Micrographs." IEEE Transactions on Applied Superconductivity, vol. 3, no. 5, 2023, pp. 2023-2030, https://doi.org/10.1109/TASC.2023.3250165
24 Hyuga, H., et al. "Friction and Wear Properties of Si3N4/Carbon Fiber Composites with Aligned Microstructure." J. Am. Ceram. Soc., vol. 88, no. 5, 2005, pp. 1239–1243, https://doi.org/10.1111/j.1551-2916.2005.00283.x
25 Song, Y., et al. "Non-Amontons frictional behaviors of grain boundaries at layered material interfaces." Nature Communications, vol. 15, no. 9487, 2024, https://doi.org/10.1038/s41467-024-53581-y
26 Hilou, E., et al. "Characterizing the spatiotemporal evolution of paramagnetic colloids in time-varying magnetic fields with Minkowski functionals." Soft Matter, vol. 16, 2020, pp. 8799–8805,

# References

https://doi.org/10.1039/D0SM01100B

27  Armstrong, R.T., et al. "Porous media characterization using Minkowski functionals: theories, applications and future directions." Transp. Porous Med, vol. 130, 2019, pp. 305–335, https://doi.org/10.1007/s11242-018-1201-4
28  Nečas, D. and Klapetek, P. "Gwyddion: an open-source software for SPM data analysis." Central European Journal of Physics, vol. 10, 2012, pp. 181–188, https://doi.org/10.2478/s11534-011-0096-2
29  Szczupak, A., Cios, G., & Jany, B. R. (2025). Collected Experimental Data For "Colorimetry and Tribology of Ultrapure Copper Surface Micromodification" [Data set]. Zenodo. https://doi.org/10.5281/zenodo.17287173

# Supporting Information

## Colorimetry and Tribology of Ultrapure Copper Surface Micromodification

Aleksandra Szczupak[1], Grzegorz Cios[2], Benedykt R. Jany[1*]

[1]Marian Smoluchowski Institute of Physics, Faculty of Physics, Astronomy and Applied Computer Science, Jagiellonian University, ul. prof. Stanisława Lojasiewicza 11, 30-348 Krakow, Poland

[2]Academic Centre for Materials and Nanotechnology, AGH University of Krakow, al. A Mickiewicza 30, 30-059 Krakow, Poland

*corresponding author e-mail: benedykt.jany@uj.edu.pl

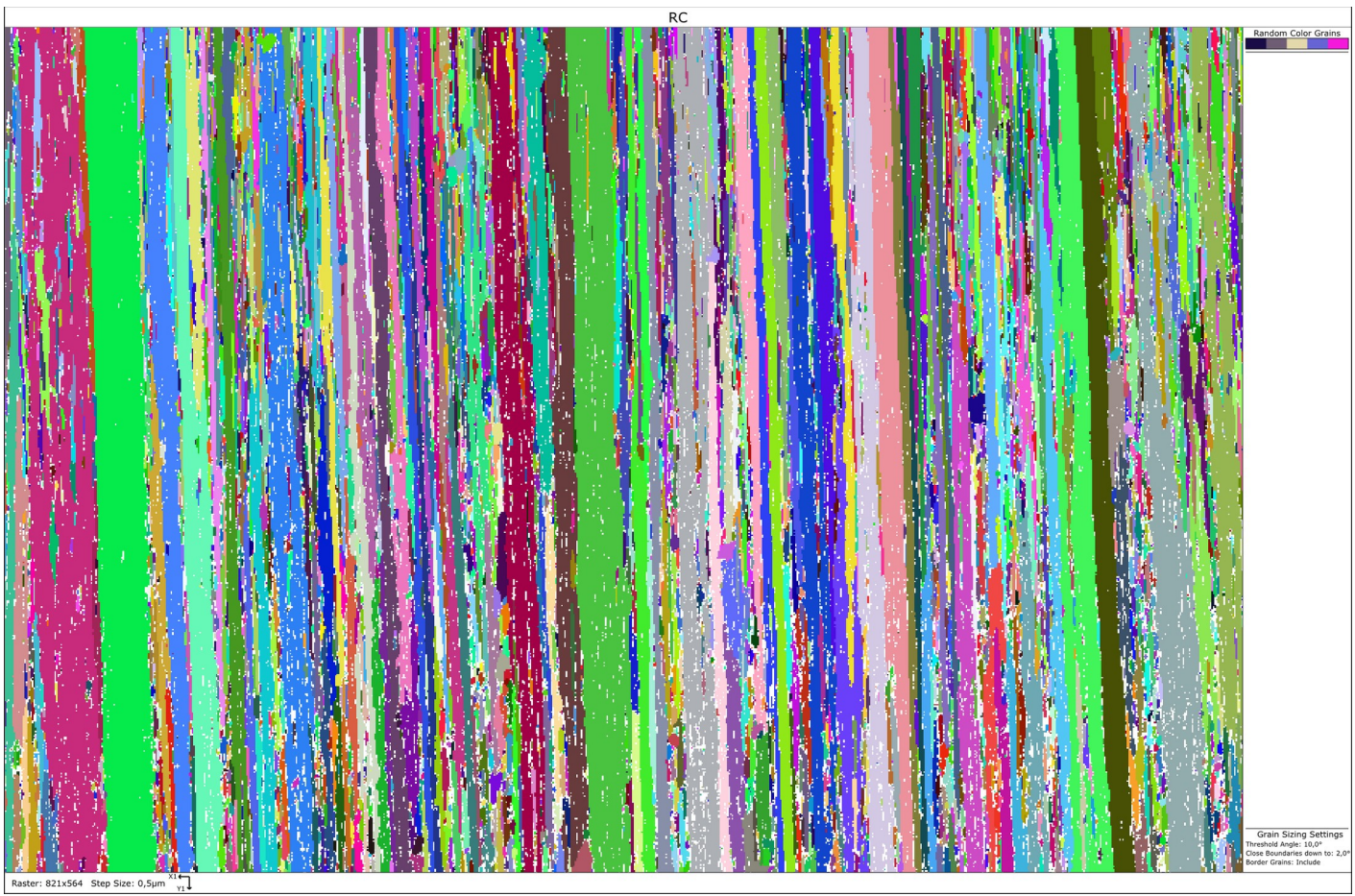

Figure S1: Grains SEM EBSD map of the ultrapure copper pellet for the misorientation angle 10 degrees.

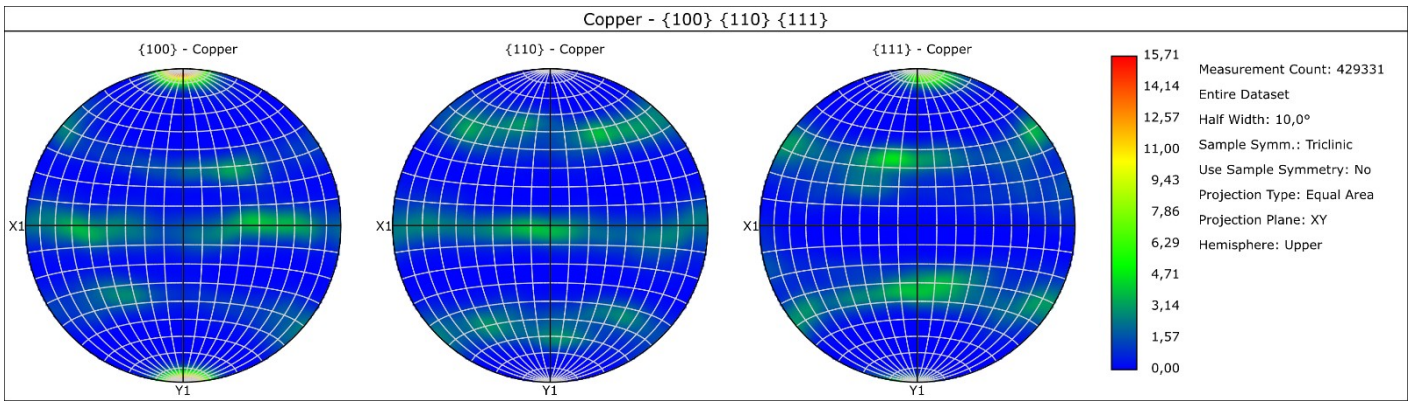

Figure S2: Pole Figures map of the ultrapure copper pellet.

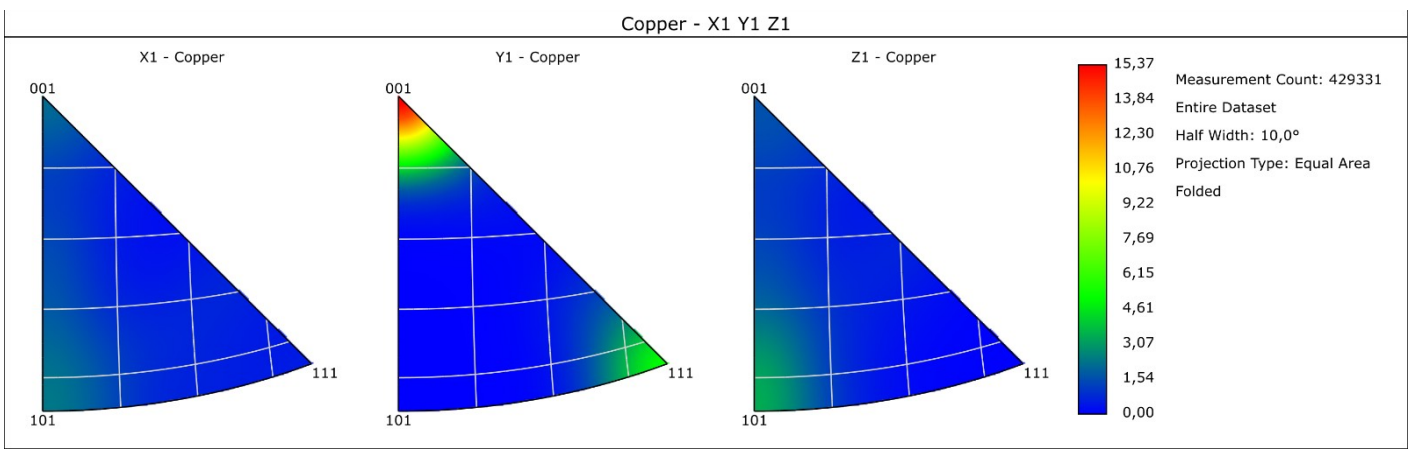

Figure S3: Inverse Pole Figures map of the ultrapure copper pellet.

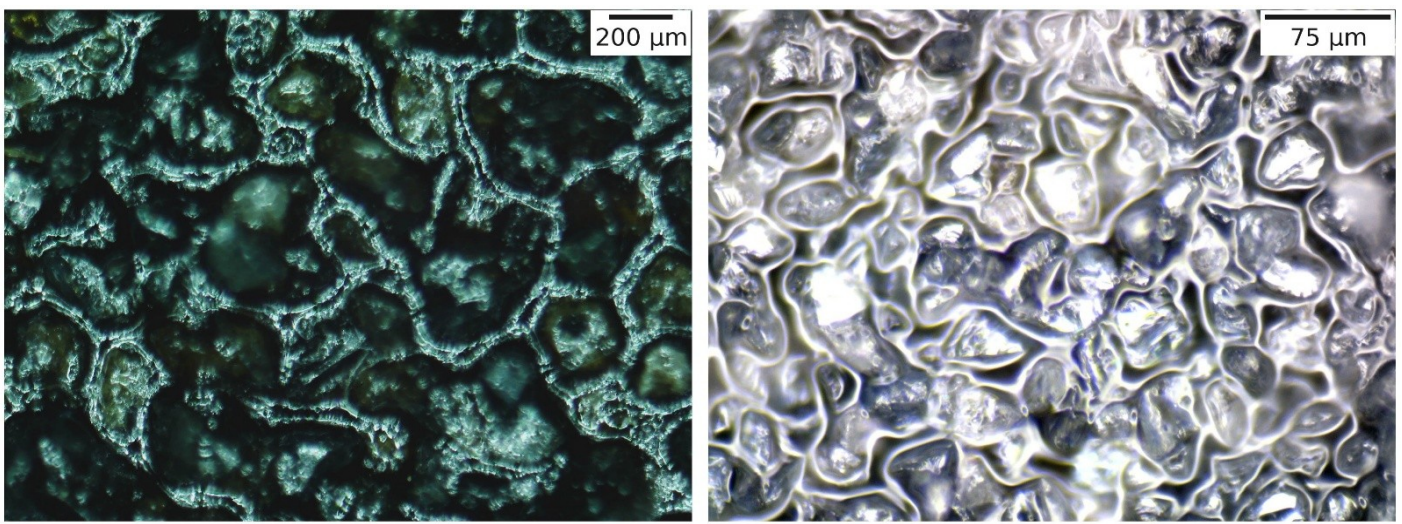

Figure S4: Example images of used sandpapers – P120 and P600.

| Paper P | Diameter $d_p$ [µm] | Distances between grains $o_p$ [µm] | Contact diameter $d_a$ [µm] | Depth of max shear stress in copper $z_{max,2}$ [µm] |
|---|---|---|---|---|
| 120 | 241.9 ± 9.7 | 292 ± 13 | 24.23 ± 0.41 | 5.864 ± 0.098 |
| 150 | 164.9 ± 7.6 | 206 ± 12 | 21.33 ± 0.39 | 5.155 ± 0.094 |
| 320 | 59.5 ± 2.4 | 88.0 ± 3.7 | 15.18 ± 0.27 | 3.679 ± 0.064 |
| 600 | 26.4 ± 1.2 | 36.1 ± 1.6 | 11.57 ± 0.21 | 2.803 ± 0.051 |
| 800 | 18.36 ± 0.95 | 25.0 ± 1.1 | 10.26 ± 0.21 | 2.477 ± 0.050 |
| 1200 | 12.88 ± 0.62 | 18.00 ± 0.62 | 9.12 ± 0.18 | 2.202 ± 0.042 |
| 1500 | 11.15 ± 0.56 | 16.12 ± 0.72 | 8.69 ± 0.17 | 2.102 ± 0.041 |
| 2000 | 10.40 ± 0.43 | 14.03 ± 0.61 | 8.49 ± 0.15 | 2.052 ± 0.035 |

Table S1: Characterization of used sandpapers: grains diameters $d_p$, distances between grains $o_p$, contact diameters with copper surface $d_a$ calculated from Hertzian Contact Model, depths of maximum shear stress in copper $z_{max,2}$ calculated from Hertzian Contact Model.

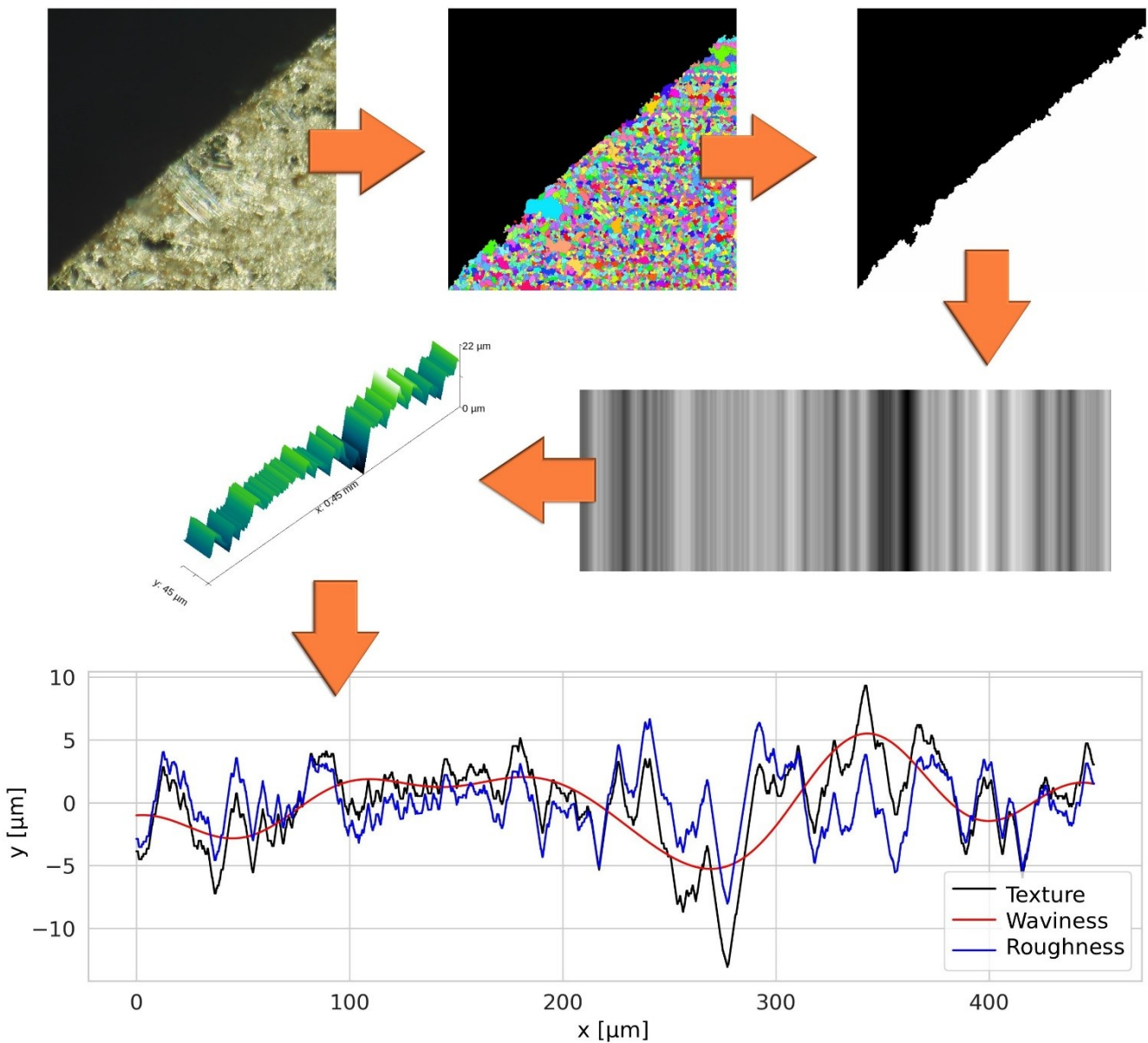

Figure S5: Schematic diagram of the roughness profile extraction method.

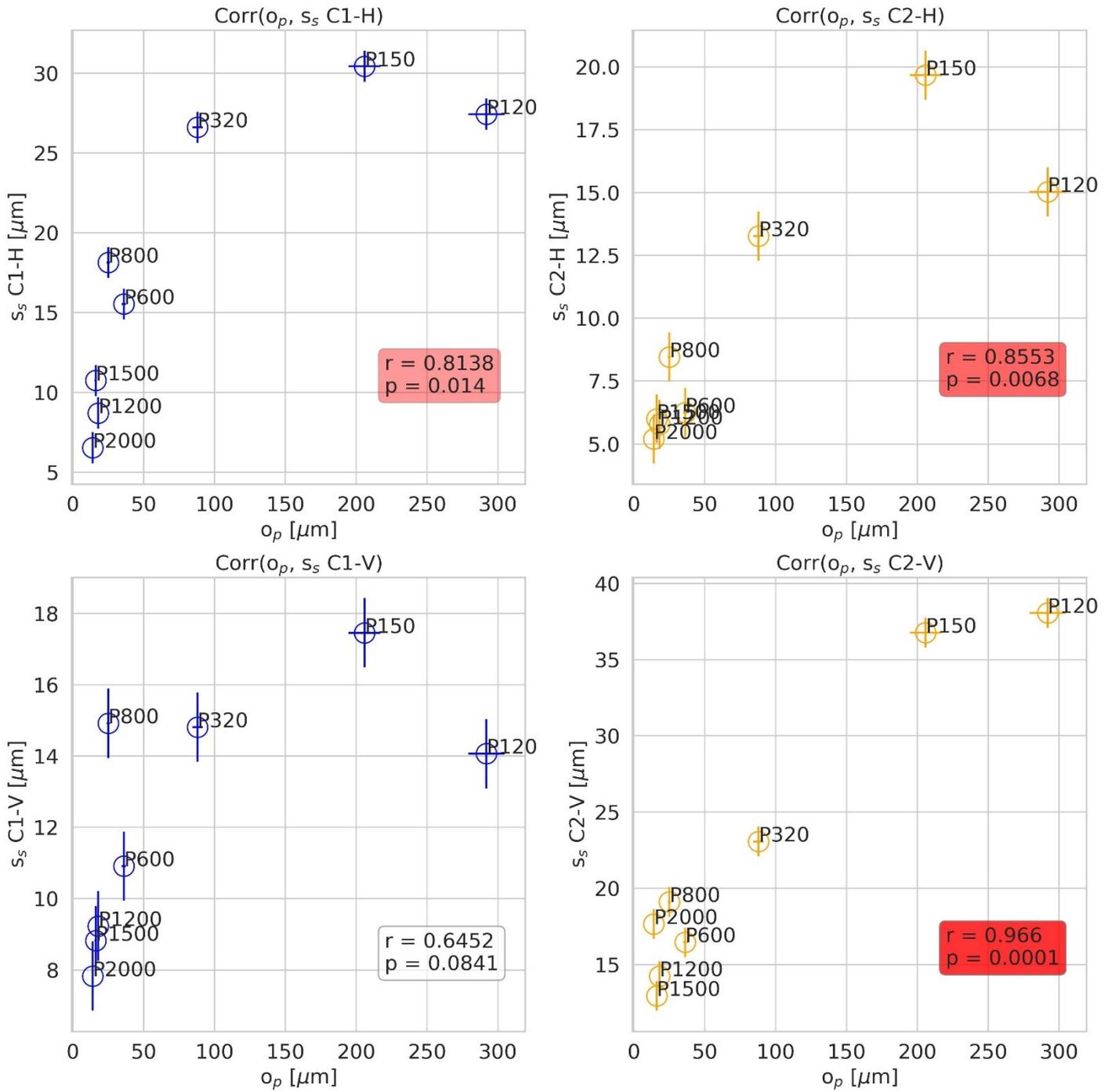

Figure S6: Relations between mean texture structure sizes $s_s$ in directions H, V for opposite micromadificaton directions C1, C2 and distances between paper grains $o_p$. Strong, statistically significant correlations for directions C2-H and C2-V.

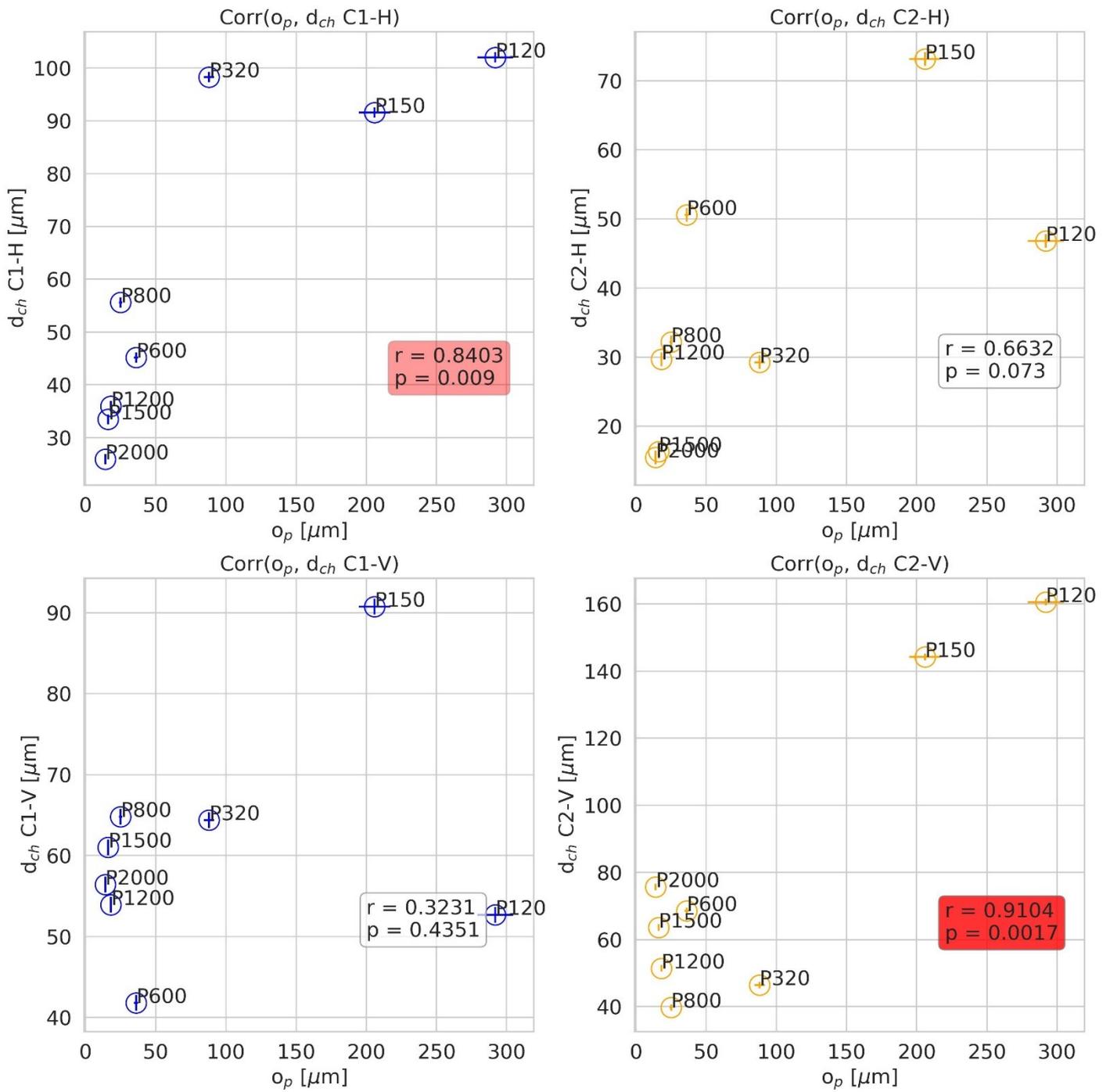

Figure S7: Relations between mean distances between texture structures $d_{ch}$ in directions H, V for opposite directions of micromodification C1, C2 and distances between paper grains $o_p$. Strong, statistically significant correlations for directions C1-H and C2-V.

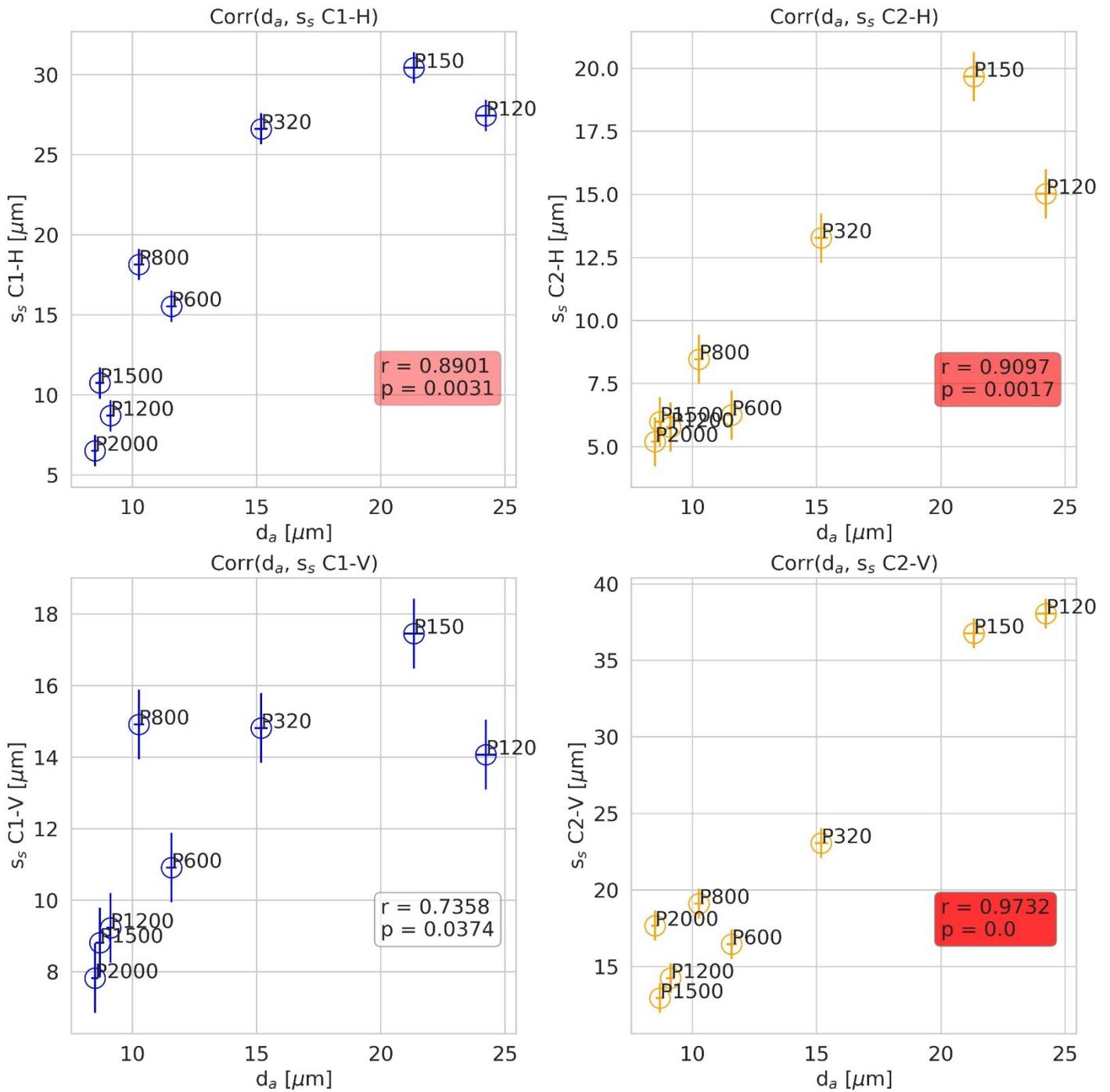

Figure S8: Reltions between mean texture structure sizes $s_s$ in the H, V directions for the opposite micromodification directions C1, C2 and contact diameters of the papers with the copper surface $d_a$. Strong, statistically significant correlations, especially for the C2-V direction.

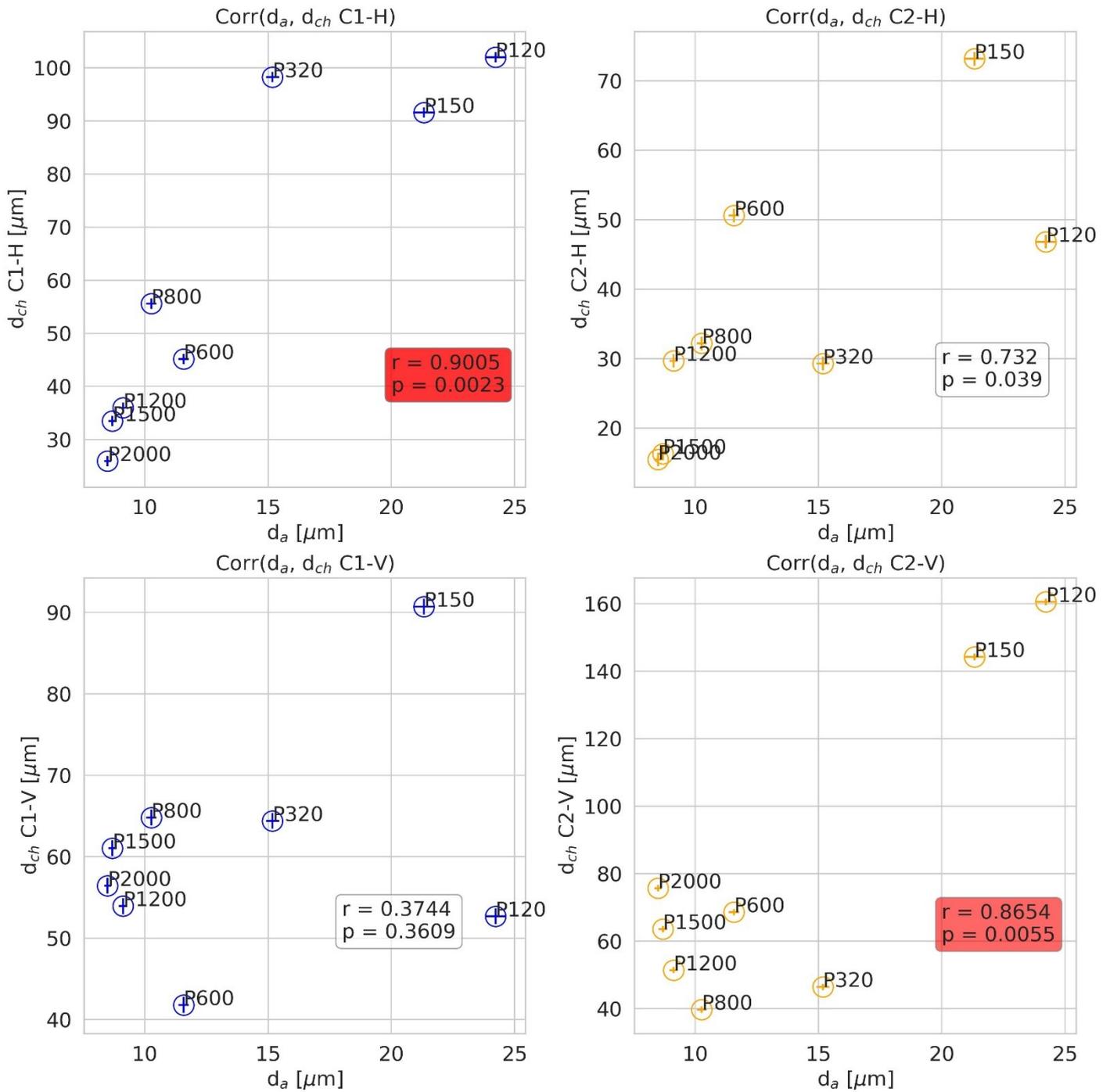

Figure S9: Relations between mean distances between texture structures $d_{ch}$ in directions H, V for opposite directions of micromodification C1, C2 and contact diameters of papers with the copper surface $d_a$. Strong, statistically significant correlations, especially for C1-H, C2-V.

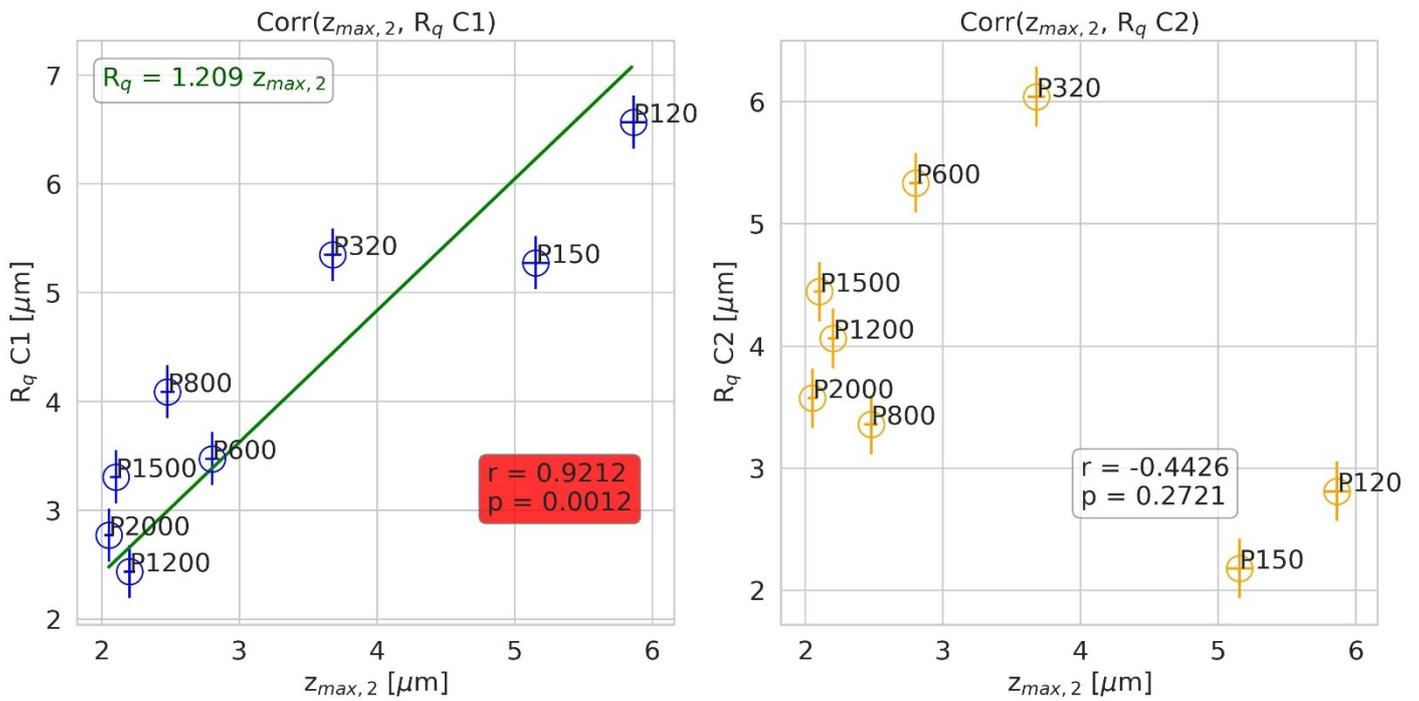

Figure S10: Relations between surface roughness $R_q$ for opposite micromodification directions C1, C2 and depths of maximum shear stress in copper $z_{max,2}$. Very strong and statistically significant correlations for the micromodification direction C1.

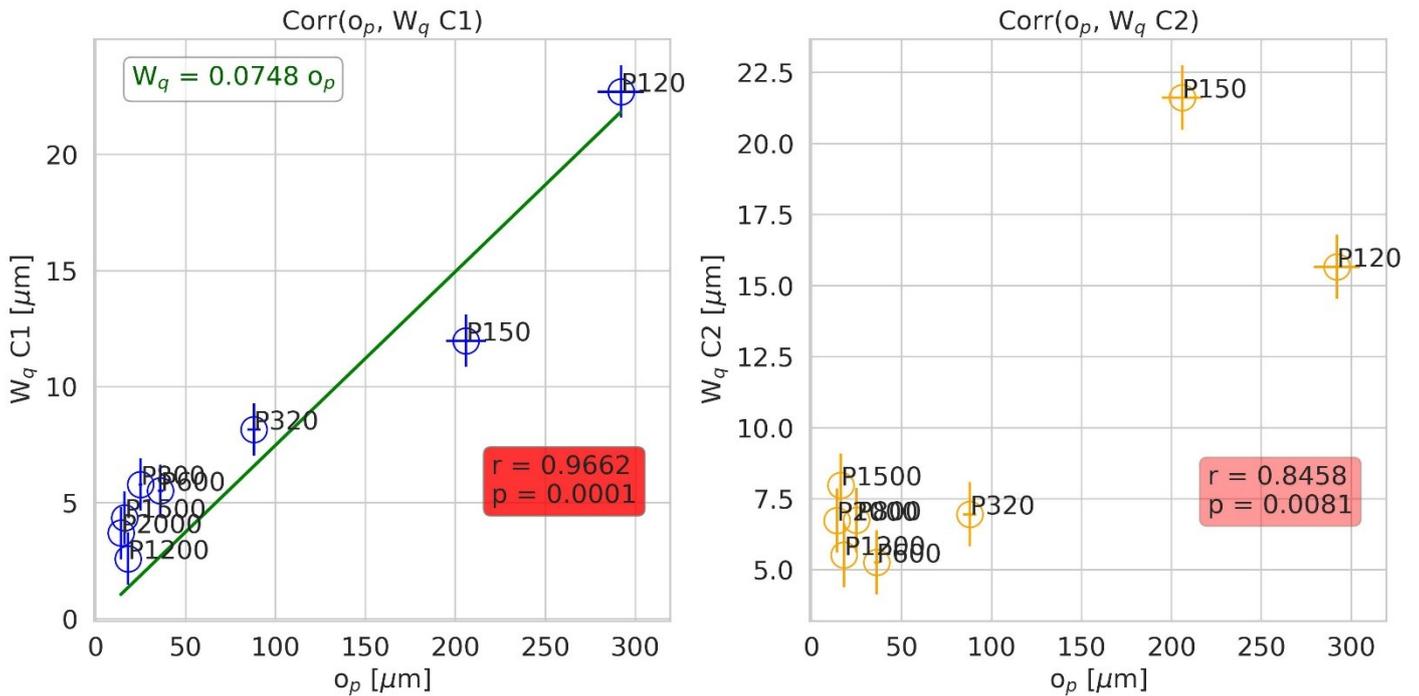

Figure S11: Relations between surface waviness $W_q$ for opposite directions of micromodification C1, C2 and distances between paper grains $o_p$. Very strong and statistically significant correlations, especially for direction C1.

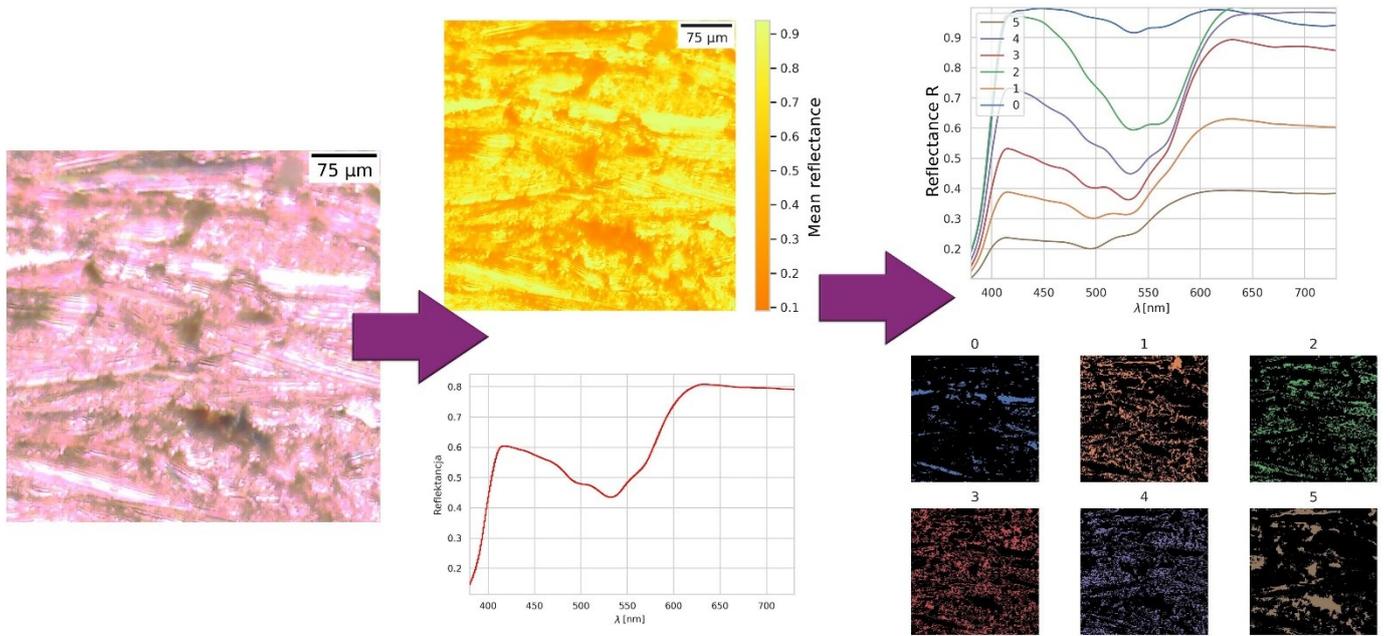

Figure S12: Scheme of hyperspectral reflectance reconstruction and clustering of hyperspectral data.

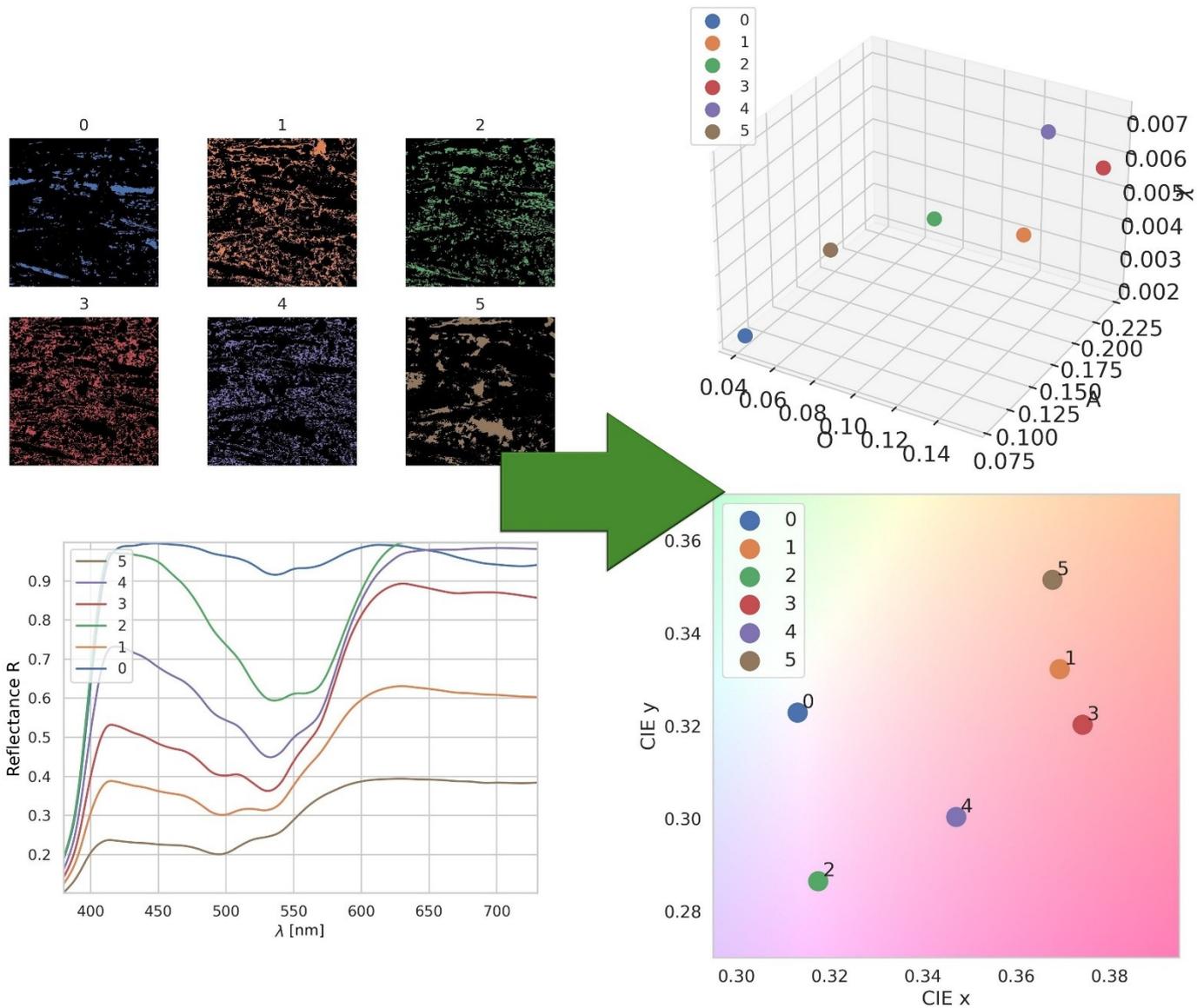

Figure S13: Scheme of the analysis of Minkowski functionals and mean color on the CIE 1931 chromaticity diagram.

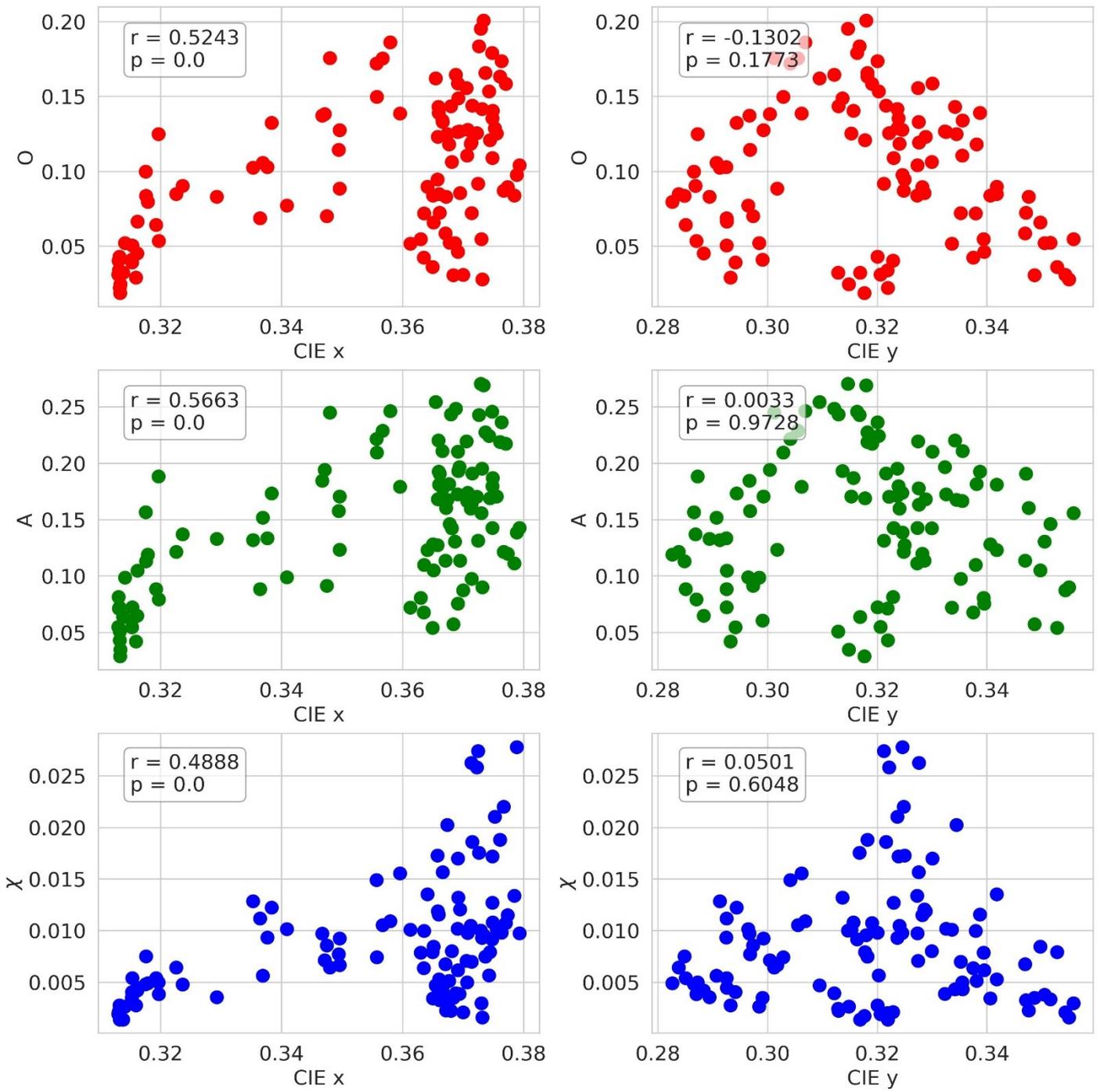

Figure S14: Correlations between morphological and optical properties of the 109 obtaines surface structures.

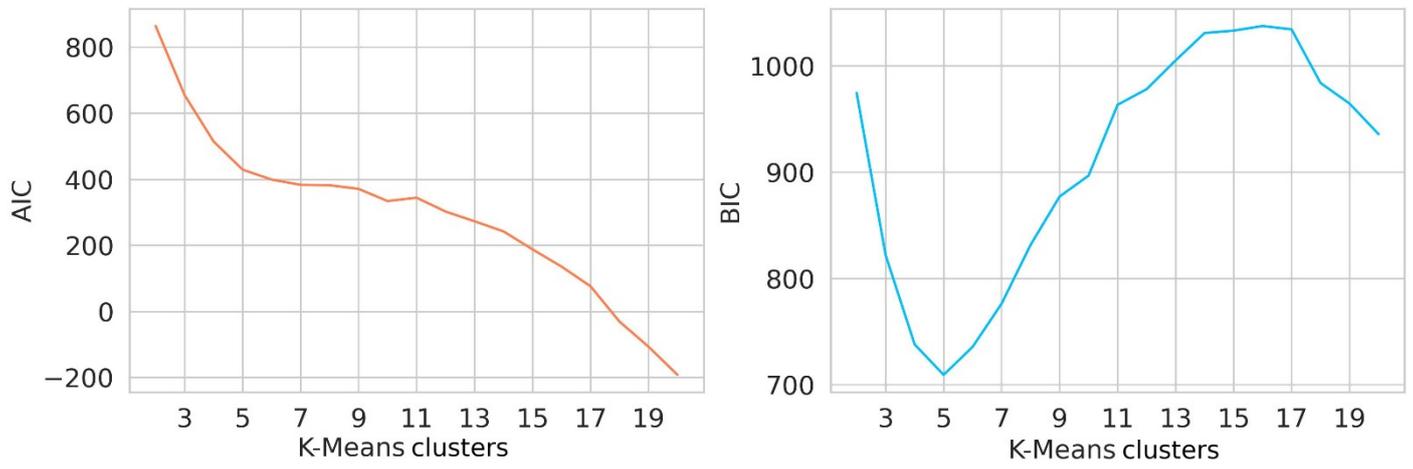

Figure S15: Information criteria for selecting models AIC and BIC used to determine the optimal number of K-Means clusters – 5 clusters.

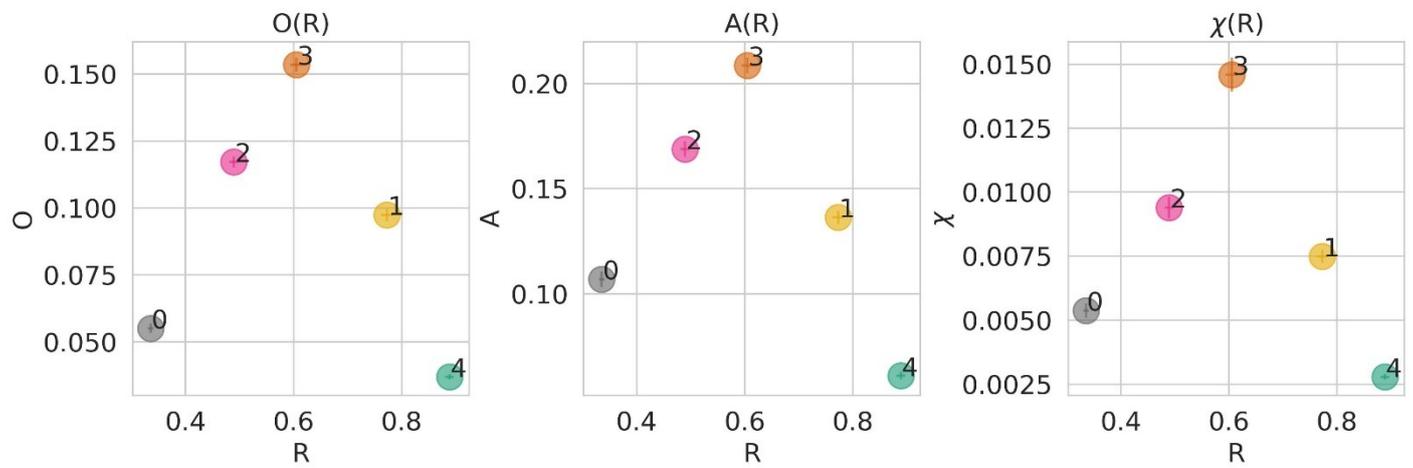

Figure S16: Relations between Minkowski functionals O, A, χ, and the mean reflectance R for clusters.

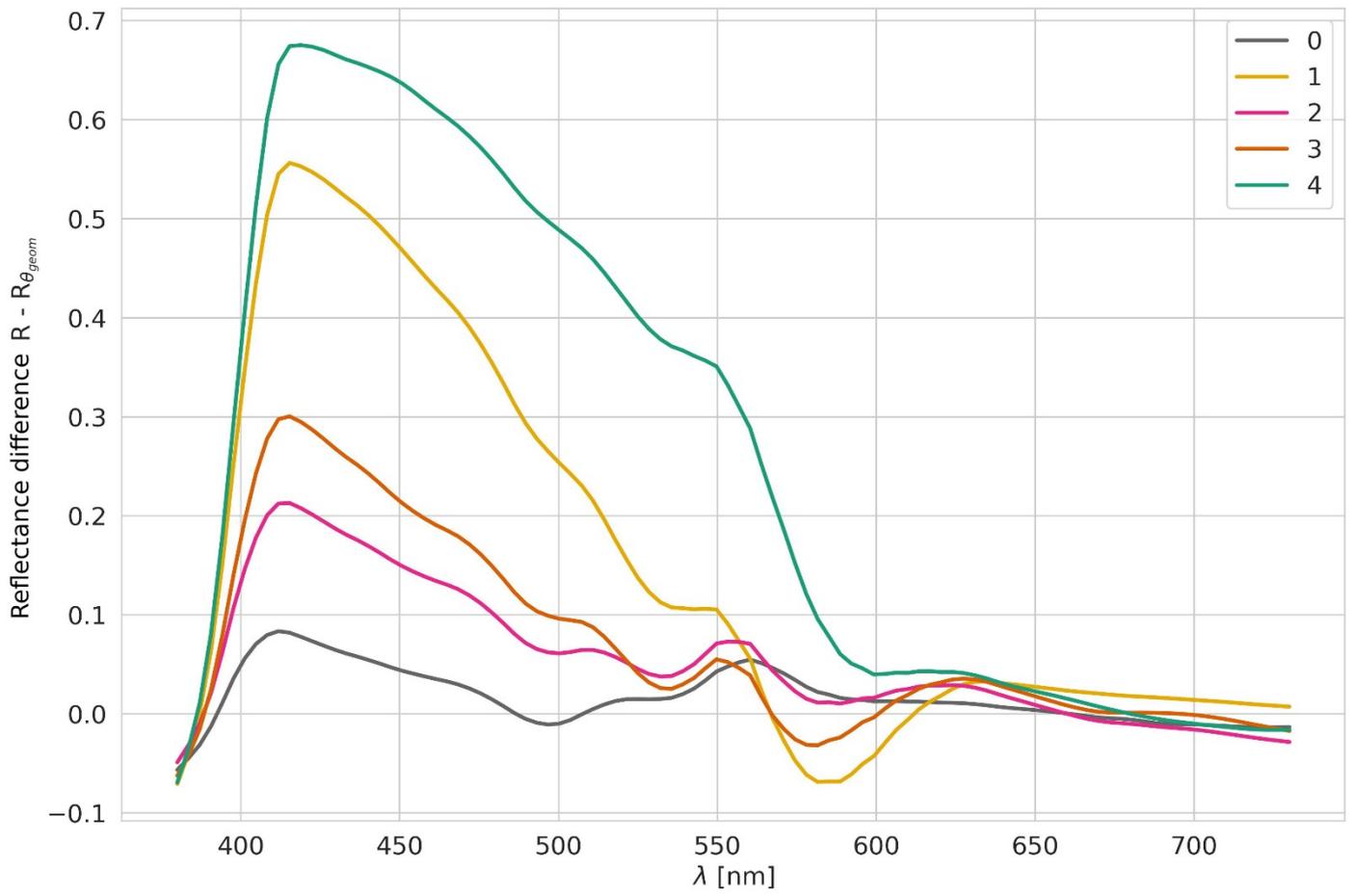

Figure S17: Differences of experimental reflectances R and reflectances calculated from the multiple reflections in microgrooves model normalized to the maximum of the experimental reflectances R.

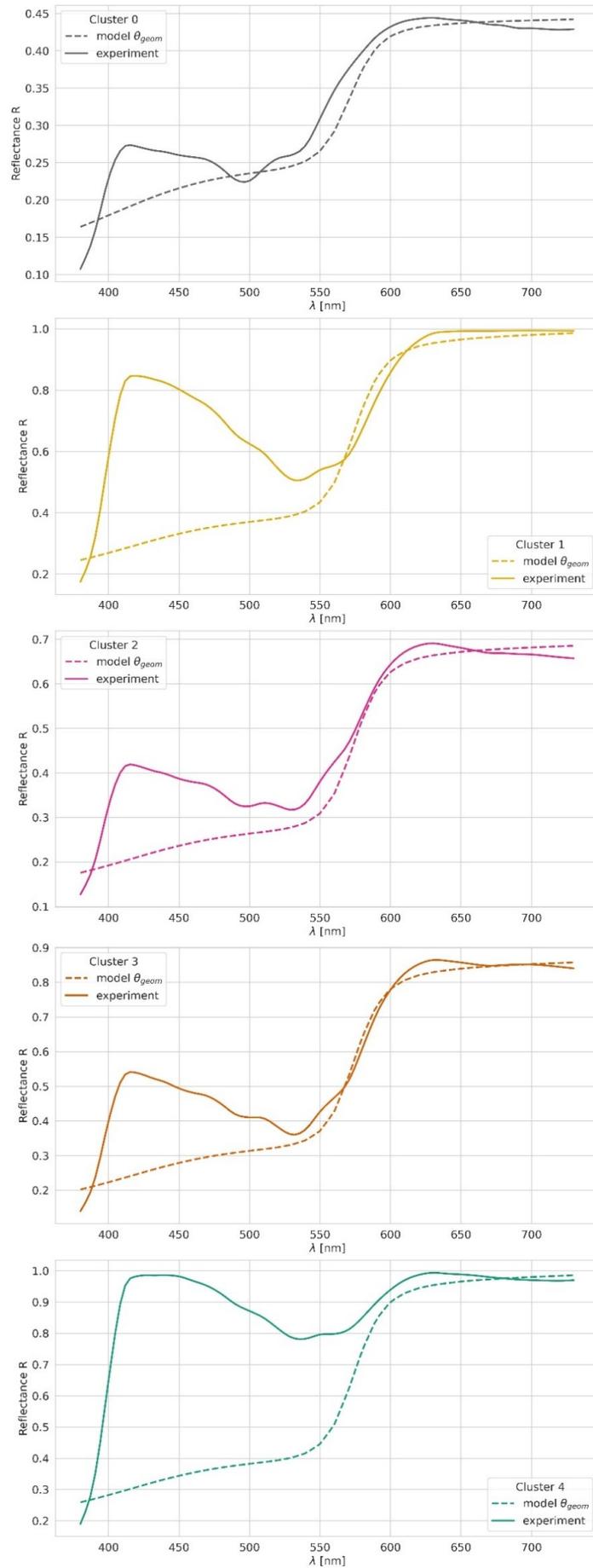

Figure S18: Comparison of experimental and model-calculated reflectances.

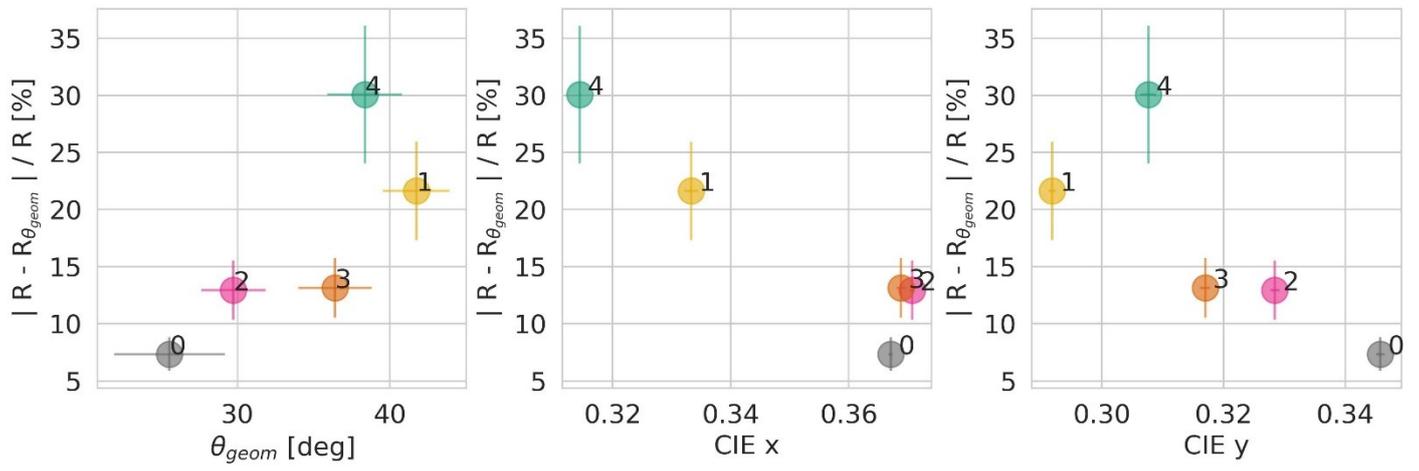

Figure S19: Dependence of the percentage value of the difference between the experimental reflectance and the reflectance calculated from the model on the inclination angles $\theta_{geom}$ and the CIE x and CIE y chromatic coordinates for clusters.